\documentclass[conference]{IEEEtran}
\IEEEoverridecommandlockouts

\usepackage{cite}
\usepackage{amsmath,amssymb,amsfonts}
\usepackage{algorithmic}
\usepackage{graphicx}
\usepackage{textcomp}
\usepackage{xcolor}
\usepackage{bbm}
\usepackage{multirow}
\def\BibTeX{{\rm B\kern-.05em{\sc i\kern-.025em b}
\kern-.08em T\kern-.1667em\lower.7ex\hbox{E}
\kern-.125emX}}
\begin{document}

\title{Beyond the QBER Threshold: A Temporal QBER Based Machine Learning Framework for Multi Attack Detection in BB84 QKD}

\author{
\setlength{\tabcolsep}{6pt}
\centering
\begin{tabular}{ccc}
\begin{minipage}[t]{0.31\textwidth}\centering
Isha\\
Drone Lab, IIT Mandi\\
S24041@students.iitmandi.ac.in
\end{minipage}
&
\begin{minipage}[t]{0.31\textwidth}\centering
Deepak Singh\\
Drone Lab, IIT Mandi\\
DD24021@students.iitmandi.ac.in
\end{minipage}
&
\begin{minipage}[t]{0.31\textwidth}\centering
Devesh Kumar\\
DRDO\\
devesh.kumar.hqr@gov.in
\end{minipage}
\\[3.5em]
\begin{minipage}[t]{0.31\textwidth}\centering
S.K Pal\\
DRDO\\
saibal.pal@gov.in
\end{minipage}
&
\begin{minipage}[t]{0.31\textwidth}\centering
Praful Hambarde\\
Drone Lab, IIT Mandi\\
praful@iitmandi.ac.in
\end{minipage}
&
\begin{minipage}[t]{0.31\textwidth}\centering
Amit Shukla\\
Drone Lab, IIT Mandi\\
amitshukla@iitmandi.ac.in
\end{minipage}
\end{tabular}
}

\maketitle

\begin{abstract}
Conventional BB84 Quantum Key Distribution (QKD) systems rely on a fixed 11\% Quantum Bit Error Rate (QBER) threshold to detect eavesdropping. However, stealthy attacks can remain below this threshold while still compromising channel security. This paper proposes a temporal QBER based machine learning framework for detecting and classifying eavesdropping attacks in BB84 QKD systems. Rather than relying on average session level QBER, the framework extracts 63 physics-informed temporal features capturing burst behavior, temporal instability, basis dependent asymmetry, and QBER loss interactions. Random Forest, XGBoost, and Support Vector Machine with a Radial Basis Function kernel (SVM-RBF) classifiers are evaluated on seven eavesdropping attacks and a normal channel scenario under noisy and lossy conditions. Averaged over ten independent runs, XGBoost achieves the best performance with 88.01\% ($\pm$0.47\%) accuracy and a macro F1 score of 0.8803, while SVM-RBF performs comparably, confirming the robustness of the proposed features. Evaluated as a binary attack-versus-normal detector for comparison with conventional monitoring, a fixed 11\% QBER threshold achieves only 25.82\% accuracy with a False Negative Rate (FNR) of 0.8477, whereas the proposed framework reduces the FNR to 0.0198, substantially improving detection of stealthy attacks that evade threshold-based monitoring. SHapley Additive exPlanations based (SHAP) explainability shows that physics-informed temporal and channel derived features are highly discriminative for identifying eavesdropping strategies. These results demonstrate that temporal QBER driven machine learning provides an accurate, explainable, and practical framework for multi attack security monitoring in BB84 QKD systems.
\end{abstract}


\section{Introduction}

Quantum Key Distribution (QKD) enables cryptographic key exchange with security 
guaranteed by the principles of quantum mechanics rather than computational 
assumptions. Among existing QKD protocols, BB84 remains one of the most widely 
adopted due to its simplicity and strong theoretical security guarantees 
\cite{bennett1984quantum}\cite{scarani2009security}. Any measurement performed by an 
eavesdropper inevitably disturbs the transmitted quantum states, increasing the 
Quantum Bit Error Rate (QBER). Consequently, practical BB84 systems commonly 
employ a fixed QBER threshold of approximately 11\% to detect potential 
eavesdropping \cite{shor2000simple}\cite {lutkenhaus2000security}.

Although effective against aggressive attacks, fixed-threshold QBER monitoring 
becomes unreliable in realistic quantum communication environments\cite{diamanti2016practical}. Practical QKD systems are affected by detector imperfections, channel noise, optical losses, and finite-key effects, which increase baseline QBER and make reliable attack detection more challenging\cite{xu2020secure}\cite{pirandola2020advances}. Under such conditions, an adversary can perform stealthy partial attacks by 
intercepting only a fraction, $r_e$, of the transmitted photons while keeping 
the induced QBER below the conventional detection threshold\cite{brassard2000limitations}. Similar limitations have also been reported for compromised random number generator attacks that evade conventional QBER-based detection\cite{saxena2025detection}.  For an intercept resend attack\cite{lizama2016quantum}, the expected disturbance approximately follows 
$\text{QBER}_{\text{IR}} \approx r_e/4$, where $r_e$ denotes the interception 
rate\cite{nielsen2010quantum}. Similar 
stealthy behavior can also arise in practical Photon Number Splitting (PNS), Beam Splitting, 
Time-Shift, Trojan Horse, Quantum 
Cloning, and Fake State attacks, allowing significant information leakage while 
remaining difficult to detect using threshold based monitoring alone\cite{cao2022evolution}\cite{mehic2020quantum}. Although decoy-state BB84 and Measurement-Device-Independent (MDI) QKD mitigate several known attacks, implementation imperfections and side-channel vulnerabilities still motivate complementary monitoring techniques based on post-processing statistics. The proposed framework acts as a complementary software-based monitoring layer that enhances practical QKD security without replacing the information-theoretic security guarantees of BB84. Beyond fixed threshold approaches, sequential change detection methods such as Cumulative Sum (CUSUM) \cite{page1954continuous}, Bayesian Online Change Point Detection 
(BOCD) \cite{adams2007bayesian}, and related quantum variants
\cite{fanizza2023ultimate} provide efficient online anomaly detection but primarily perform binary change detection rather than multiclass attack identification. Unlike binary detection, multiclass attack identification provides actionable information for effective mitigation and incident response in practical QKD systems.

Recent studies have explored machine learning techniques for QKD security monitoring \cite{banerjee2025machine} using Random Forests \cite{ding2023machine}\cite{tunc2023machine}, Support Vector Machines (SVMs)\cite{cortes1995support}, deep learning models, hybrid quantum classical approaches\cite{al2026resisting}, and supervised learning for QBER-based quantum noise classification.Although these approaches have demonstrated promising performance, they typically focus on binary detection, limited attack scenarios, or scalar-QBER features\cite{al2026machine}.

To address these limitations, this paper proposes a temporal-QBER-based machine learning framework for detecting and classifying eavesdropping attacks in BB84 QKD systems. Rather than relying solely on average session-level QBER, the proposed framework extracts 63 physics-informed statistical, temporal, basis-dependent, and channel-aware features from standard BB84 post-processing information without requiring additional hardware modifications. The framework complements standard BB84 post-processing without modifying existing protocols. These features enable the detection of subtle attack signatures that are difficult to identify using scalar QBER or binary change detection alone. Unlike existing studies that focus on binary attack detection, the proposed framework performs fine-grained multiclass classification to identify specific eavesdropping strategies while remaining compatible with future online CUSUM-based monitoring.
Random Forest\cite{breiman2001random}, XGBoost\cite{chen2016xgboost}, and SVM-RBF are evaluated on seven practical eavesdropping attacks and a normal channel under noisy and lossy communication conditions. Experimental results averaged over ten 
independent runs demonstrate that XGBoost achieves the best overall performance 
with an average classification accuracy of 88.01\% ($\pm$0.47\%) and a macro-F1 
score of 0.8803, while SVM-RBF achieves comparable performance. When evaluated as a binary detector, the conventional fixed 11\% QBER threshold achieves only 25.82\% accuracy with a False Negative Rate (FNR) of 0.8477, whereas the proposed framework reduces the FNR to 0.0198, substantially improving the detection of stealthy attacks. The main contributions of this work are summarized as follows:

\begin{itemize}
   \item  A temporal-QBER feature engineering framework com-
prising 63 physics-informed features for multi-class
eavesdropping attack detection and classification in BB84
QKD systems.
\item A realistic BB84 simulation framework supporting seven
practical eavesdropping attacks under noisy and lossy
channel conditions.
\item A comprehensive comparative evaluation of Random Forest, XGBoost, and SVM-RBF classifiers
for multi attack QKD security monitoring, together with
a comparison against the conventional fixed-threshold
QBER detector.
\item  A security-oriented evaluation emphasizing the FNR alongside conventional classification metrics.
\item SHAP-based explainability analysis of the proposed temporal and channel-aware features.
\end{itemize}

The remainder of this paper is organized as follows. Section II presents the proposed Temporal-QBER Detection framework. Section III describes the experimental setup and performance evaluation. Section IV discusses the results and future research directions. Finally, Section V concludes the paper.

\section{Proposed Temporal-QBER Detection Framework}
\begin{figure}[t]
\centering
\includegraphics[width=0.95\columnwidth]{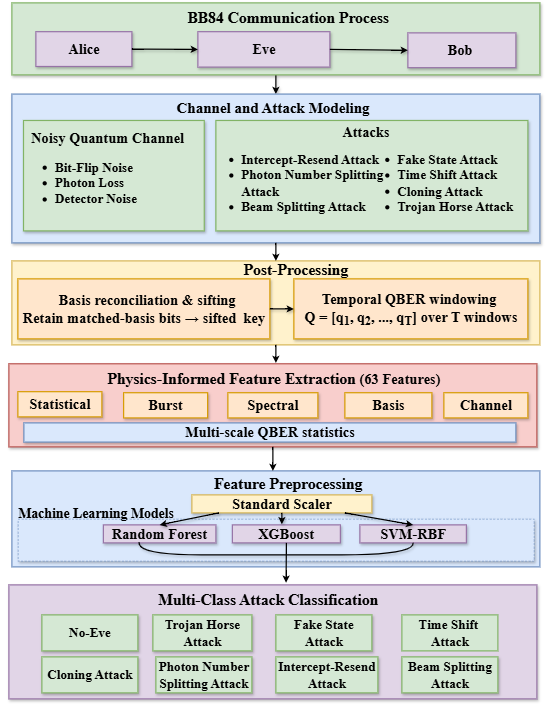}
\caption{Proposed temporal QBER based machine learning framework for multi-attack detection in BB84 QKD systems.} \label{fig:framework}
\end{figure}

The proposed framework performs intelligent security monitoring for BB84 QKD systems using temporal-QBER analysis and machine learning. As shown in Fig.~\ref{fig:framework}, the framework generates temporal QBER sequences, extracts 63 physics-informed features, and classifies normal communication and seven eavesdropping attacks using machine learning classifiers.

\subsection{BB84 Communication and Threat Model}

The proposed framework considers a standard BB84 QKD system in which Alice prepares randomly polarized photons and Bob performs measurements using independently selected bases. After basis reconciliation, only matching measurement outcomes are retained to form the sifted key. The communication quality is quantified using the QBER, defined as

\begin{equation}
\mathrm{QBER}=\frac{1}{n}\sum_{i=1}^{n}(a_i\oplus b_i)
\end{equation}

where $a_i$ and $b_i$ denote Alice's transmitted bit and Bob's received sifted bit, respectively, and $n$ is the sifted-key length.

The threat model considers seven practical eavesdropping attacks namely Intercept-Resend, Photon Number Splitting, Beam Splitting, Time-Shift, Trojan Horse, Quantum Cloning, and Fake State. These attacks produce distinct temporal QBER patterns through different combinations of burst errors, basis-dependent disturbances, and photon-loss characteristics. To reduce simulation label leakage, overlapping noise and loss distributions are maintained across normal and attack sessions. The resulting temporal QBER sequences form the basis of the proposed feature extraction framework described in the following subsection.

\subsection{Temporal-QBER Monitoring Framework}

Instead of relying solely on average session level QBER, the proposed framework analyzes temporal QBER variations throughout each BB84 communication session. The sifted key is represented as

\begin{equation}
S=\{(a_i,b_i)\}_{i=1}^{N}
\end{equation}

where $a_i$ and $b_i$ denote Alice's transmitted bit and Bob's measured bit, respectively, and $N$ is the sifted-key length. The sifted key is divided into non-overlapping windows of size $W$, producing

\begin{equation}
T=\left\lfloor\frac{N}{W}\right\rfloor
\end{equation}

temporal segments. For each window, the local QBER is computed as

\begin{equation}
q_t=\frac{1}{W}\sum_{i=(t-1)W+1}^{tW}(a_i\oplus b_i)
\end{equation}

yielding the temporal-QBER sequence $Q=[q_1,q_2,\ldots,q_T]$. The mean and variance of the sequence are

\begin{equation}
\mu_q=\frac{1}{T}\sum_{t=1}^{T}q_t,
\end{equation}

\begin{equation}
\sigma_q^2=\frac{1}{T}\sum_{t=1}^{T}(q_t-\mu_q)^2
\end{equation}

respectively. To quantify abrupt temporal changes, the fluctuation energy is defined as

\begin{equation}
E_f=\sum_{t=2}^{T}(q_t-q_{t-1})^2
\end{equation}

where larger values indicate stronger burst like transitions. Frequency domain characteristics are obtained from the mean centered temporal-QBER sequence using the discrete Fourier transform (DFT),

\begin{equation}
F(m)=\sum_{t=1}^{T}(q_t-\mu_q)e^{-j2\pi mt/T}
\end{equation}

whose magnitude spectrum is used to derive spectral features. This representation captures burst behavior, temporal correlations, basis-dependent instability, and channel variations hidden by average-QBER monitoring. All features are derived from standard BB84 post-processing without additional hardware.

\subsection{Temporal Feature Extraction}

A total of 63 physics-informed features are extracted from the temporal-QBER sequence and grouped into five complementary categories Table~\ref{tab:features} summarizes representative features from each category. Together, these feature groups capture complementary statistical, temporal, spectral, basis-dependent, and channel-level characteristics, enabling robust discrimination of diverse eavesdropping behaviors. Temporal dependency is quantified using the lag-$k$ autocorrelation coefficient

\begin{table}[t]
\caption{Physics-Informed Temporal-QBER Feature Categories Used for Multi-Attack Detection}
\label{tab:features}
\centering
\scriptsize
\renewcommand{\arraystretch}{1.15}
\begin{tabular}{|p{2.2cm}|p{3.8cm}|p{1.7cm}|}
\hline
\textbf{Feature Category} & \textbf{Representative Features} & \textbf{Purpose} \\
\hline

Statistical Features &
Mean, variance, standard deviation, skewness, kurtosis, IQR, coefficient of variation &
Overall QBER statistics \\
\hline

Burst \& Instability Features &
Spike density, burst duration, jump energy, temporal drift, error-gap statistics &
Burst and temporal changes \\
\hline

Spectral \& Temporal Features &
Autocorrelation, DFT energy, spectral entropy, Shannon entropy &
Frequency and temporal patterns \\
\hline

Basis-Dependent Features &
Rectilinear QBER, diagonal QBER, basis asymmetry, basis burst energy &
Basis-specific disturbances \\
\hline

Channel Interaction Features &
Key generation rate, transmission efficiency, QBER--loss interaction, multi-scale QBER statistics &
Channel and loss behavior \\
\hline

\end{tabular}
\end{table}
\begin{equation}
R(k)=
\frac{\sum_{t=1}^{T-k}(q_t-\mu_q)(q_{t+k}-\mu_q)}
{\sum_{t=1}^{T}(q_t-\mu_q)^2}
\end{equation}

while spectral entropy is computed as

\begin{equation}
H_s=-\sum_{i=1}^{M}p_i\log_2(p_i)
\end{equation}

where $p_i$ denotes the normalized DFT magnitude spectrum and $M$ is the number of frequency bins. The extracted features are standardized using training-set statistics before training the Random Forest, XGBoost, and SVM-RBF classifiers. The overall detection pipeline is

\begin{equation}
Q \rightarrow X \rightarrow f(X) \rightarrow \hat{y}
\label{eq:pipeline}
\end{equation}

where $Q$ denotes the temporal-QBER sequence, $X$ represents the extracted feature vector, and $\hat{y}$ denotes the predicted communication state or attack category.

\subsection{Multi-Class Attack Classification}

For each communication session, the extracted temporal-QBER features form a feature vector

\begin{equation}
X=[x_1,x_2,\ldots,x_n]
\end{equation}

where \(n=63\) denotes the total number of extracted features. A classifier \(f(\cdot)\) predicts the communication class as

\begin{equation}
\hat{y}=f(X)
\end{equation}

where \(\hat{y}\) denotes one of eight output classes corresponding to either normal BB84 communication or one of seven practical eavesdropping attacks. Three machine learning classifiers are evaluated: RF, XGBoost, and SVM-RBF. These classifiers were selected because they represent complementary ensemble- and kernel-based learning approaches. RF employs an ensemble of decision trees, XGBoost models complex feature interactions using gradient-boosted trees, and SVM-RBF performs nonlinear classification using the kernel

\begin{equation}
K(x_i,x_j)=\exp\!\left(-\gamma\|x_i-x_j\|^2\right)
\end{equation}

where $\gamma$ controls the kernel width. The eight output classes comprise one normal communication class and seven eavesdropping attacks: Intercept-Resend, Photon Number Splitting, Beam Splitting, Time-Shift, Trojan Horse, Quantum Cloning, and Fake State.

\section{EXPERIMENTAL SETUP AND RESULTS}

\subsection{Simulation Environment and Evaluation Metrics}
All experiments were implemented in Python using Scikit-learn, XGBoost, NumPy, and Pandas. A balanced dataset of 24,000 simulated BB84 communication sessions was generated across eight classes, comprising seven eavesdropping attacks and one normal communication scenario (3,000 samples per class). To reduce simulation-label leakage, overlapping detector-noise and photon-loss distributions were maintained across all classes. The sifted key was partitioned into temporal windows of $W=50$ bits, from which 63 physics-informed temporal-QBER features were extracted. A window size of $W=50$ was selected empirically, providing a suitable trade-off between temporal resolution and feature stability. An 80/20 stratified train/test split was employed, and all results are reported as the mean $\pm$ standard deviation over ten independent runs. Hyperparameter tuning was performed using 5-fold stratified cross-validation on the training set only to prevent information leakage. A conventional fixed 11\% QBER threshold detector served as the baseline. Performance was evaluated using accuracy, macro-F1 score, precision, ROC-AUC, and the FNR.

\begin{figure}[t!]
\centering
\includegraphics[width=0.95\columnwidth]{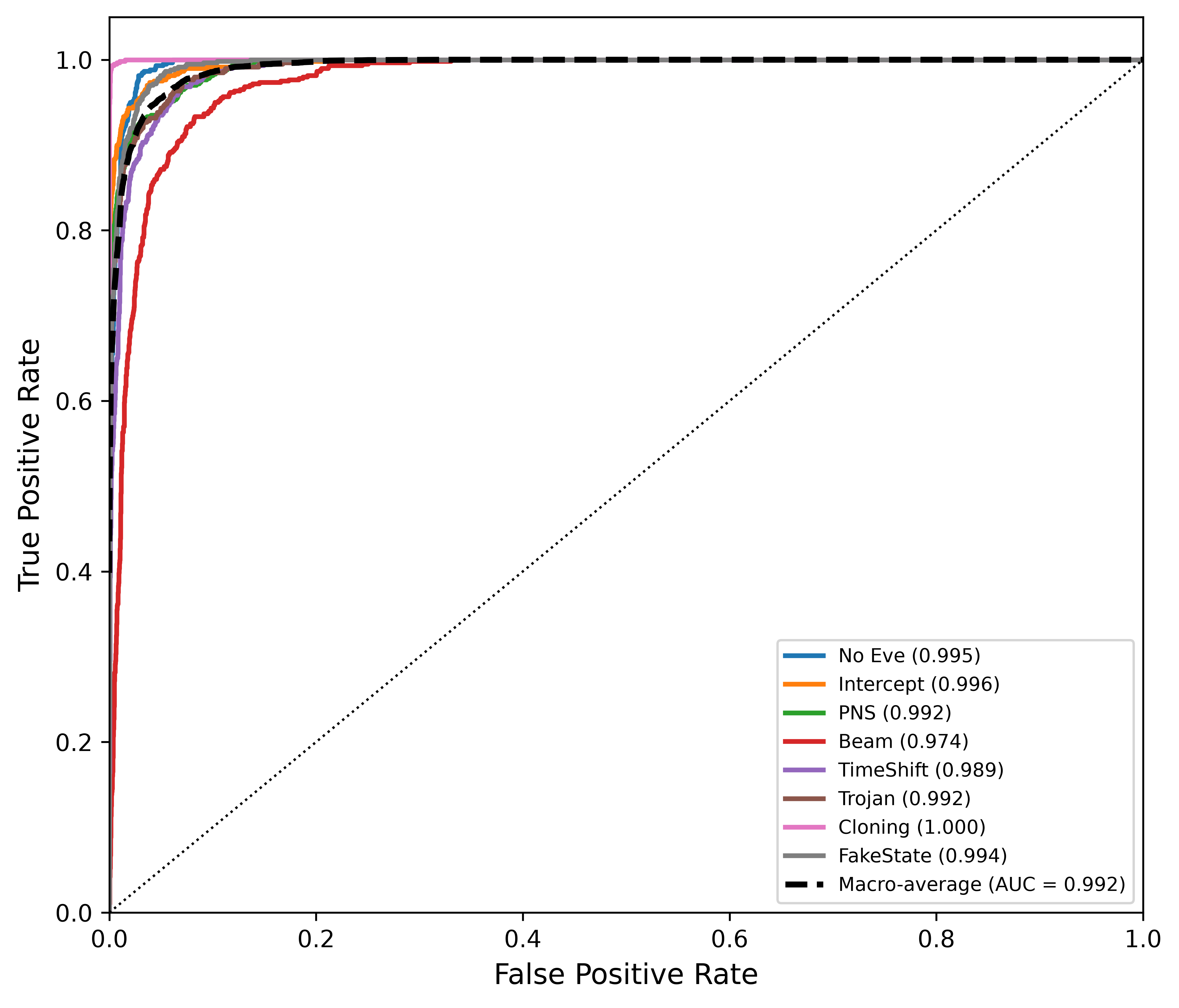}
\caption{One-vs-rest ROC curves of the XGBoost classifier for eight-class BB84 attack classification. The macro-average AUC is 0.992.}
\label{fig:roc}
\end{figure}

\begin{figure}
\centering
\includegraphics[width=0.95\columnwidth]{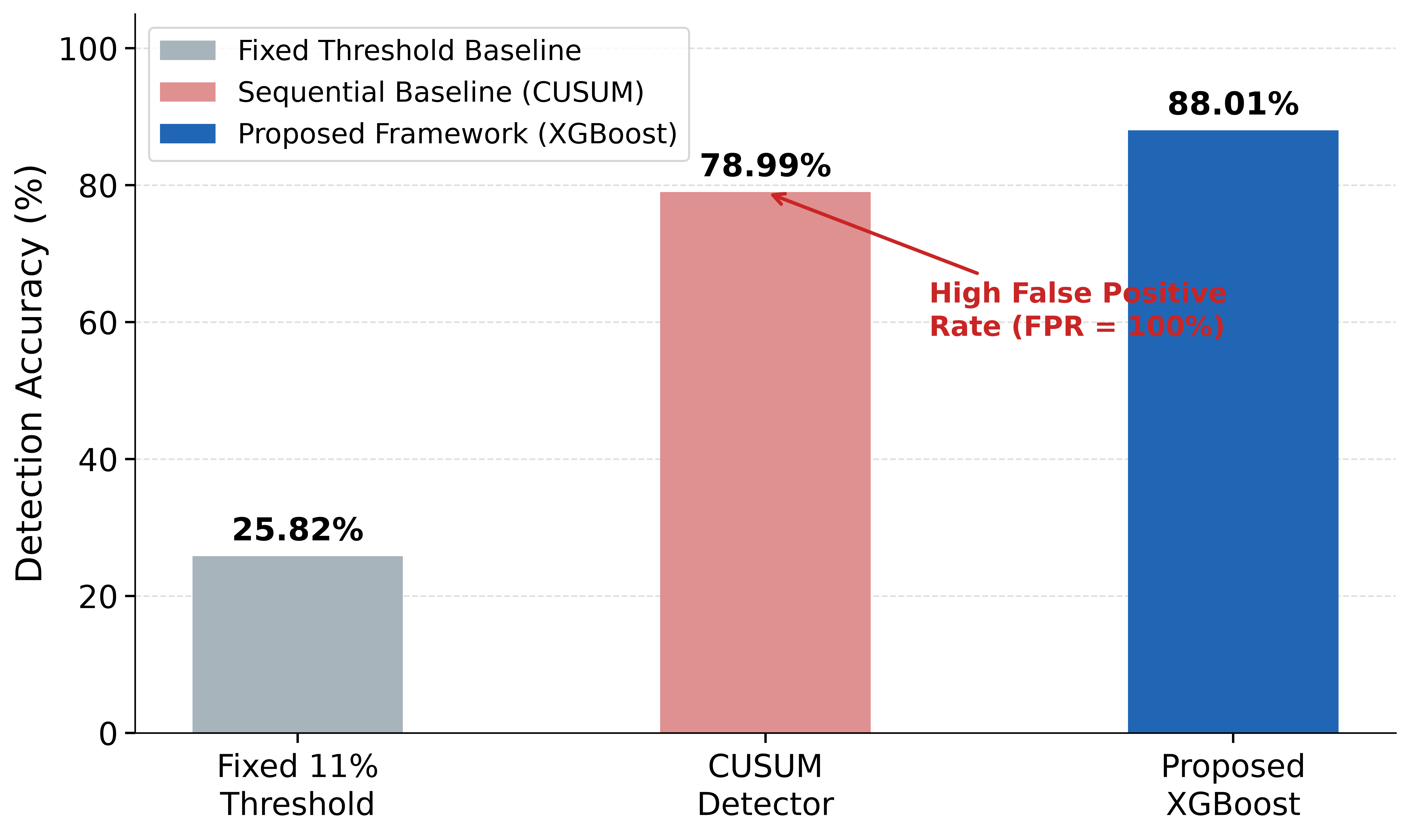}
\caption{Compares the detection accuracy of the proposed framework with conventional threshold-based and CUSUM-based monitoring approaches.}
\label{fig:baseline_comparison}
\end{figure}

\begin{figure*}
\centering
\includegraphics[width=\textwidth]{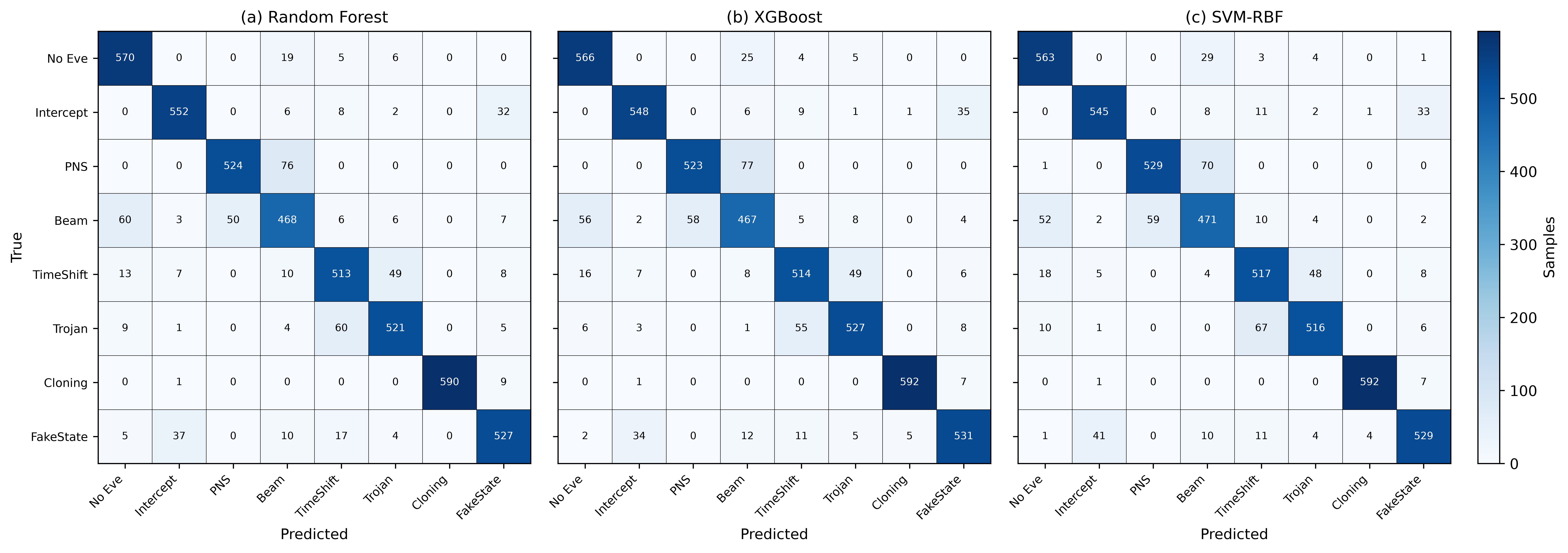}
\caption{Confusion matrices of (a) Random Forest, (b) XGBoost, and (c) SVM-RBF for eight class BB84 attack classification. XGBoost exhibits improved class separation with fewer misclassifications than the Random Forest baseline, while SVM-RBF achieves the strongest per-class discrimination. The principal confusion occurs between PNS and Beam Splitting attacks and between TimeShift and Trojan attacks due to their similar temporal-QBER characteristics.}
\label{fig:cm_compare}
\end{figure*}

\subsection{Overall Classification Performance}

Table~\ref{tab:main_results} summarizes the overall multi-class classification performance of the evaluated classifiers. XGBoost achieves the best performance with an average accuracy of 88.01\% ($\pm$0.47\%), a macro-F1 score of 0.8803, and a precision of 0.897, while SVM-RBF achieves comparable performance. The low standard deviations indicate stable and reproducible performance across repeated runs. As shown in Fig.~\ref{fig:roc}, the proposed XGBoost classifier achieves a macro-average ROC-AUC of 0.992, demonstrating excellent class separability across all eight classes.
For binary detection, Fig.~\ref{fig:baseline_comparison} compares the proposed framework with conventional monitoring approaches. The fixed 11\% QBER threshold detector achieves only 25.82\% detection accuracy with an FNR of 84.77\%. The CUSUM detector improves the detection accuracy to 78.99\% and reduces the FNR to 9.73\%, but operates with a high false-positive rate under the selected operating threshold. In comparison, the proposed XGBoost framework achieves the highest detection accuracy (88.01\%) while reducing the FNR to 1.98\%, demonstrating more reliable detection of stealthy eavesdropping attacks than conventional threshold- and CUSUM-based methods.

\begin{table}[t!]
\centering
\caption{Overall multi-class classification performance of the evaluated classifiers. Results are averaged over ten independent runs.}
\label{tab:main_results}
\resizebox{\columnwidth}{!}{%
\begin{tabular}{|l|c|c|c|}
\hline
\textbf{Model} &
\textbf{Accuracy} &
\textbf{Macro-F1} &
\textbf{Precision} \\
\hline

Random Forest &
0.757 $\pm$ 0.006 &
0.721 $\pm$ 0.006 &
0.716 \\
\hline

\textbf{XGBoost} &
\textbf{0.8801 $\pm$ 0.0047} &
\textbf{0.8803 $\pm$ 0.0047} &
\textbf{0.897} \\
\hline

SVM-RBF &
0.873 $\pm$ 0.0054 &
0.873 $\pm$ 0.0054 &
0.874 \\
\hline

\end{tabular}}
\end{table}

\subsection{Per-Class Detection Analysis}
Table~\ref{tab:perclass_f1} compares the per-class F1 scores 
of the evaluated classifiers, while Fig.~\ref{fig:cm_compare} 
presents their confusion matrices for the best-performing 
seed. XGBoost achieves the strongest per-class performance, 
with Cloning at F1 = 0.999 and consistently high scores 
across all categories. Beam Splitting yields the lowest F1 
(0.786), consistent with its lowest AUC (0.974), reflecting 
its low-amplitude attack signature. The principal confusion 
across all classifiers occurs between PNS and Beam Splitting 
and between TimeShift and Trojan, as these attack pairs 
produce similar temporal-QBER characteristics.

\begin{table}[t]
\centering
\caption{Per-class F1-score comparison of the evaluated classifiers obtained from the best-performing run.}
\label{tab:perclass_f1}
\begin{tabular}{|l|c|c|c|}
\hline
\textbf{Attack Class} & \textbf{RF} & \textbf{XGBoost} & 
\textbf{SVM-RBF} \\
\hline
No Eve     & 0.907 & 0.908 & 0.904 \\
Intercept  & 0.919 & 0.932 & 0.912 \\
PNS        & 0.893 & 0.886 & 0.891 \\
Beam       & 0.785 & 0.786 & 0.790 \\
TimeShift  & 0.849 & 0.859 & 0.848 \\
Trojan     & 0.877 & 0.883 & 0.876 \\
Cloning    & 0.992 & 0.999 & 0.989 \\
FakeState  & 0.887 & 0.923 & 0.892 \\
\hline
\end{tabular}
\end{table}

\subsection{SHAP-Based Explainability Analysis}

\begin{figure}[t]
\centering
\includegraphics[width=0.95\columnwidth]{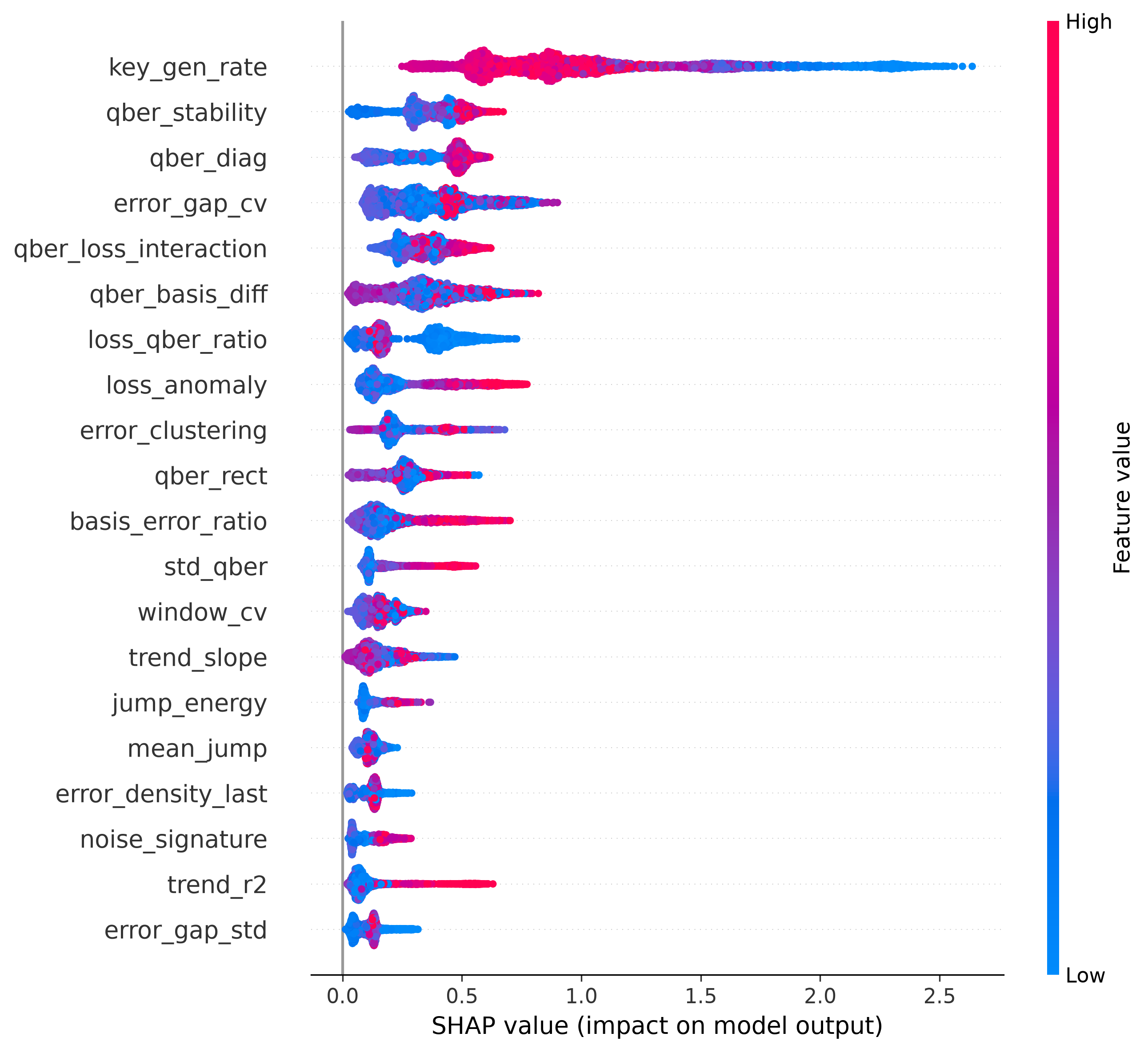}
\caption{Global SHAP feature importance of the proposed XGBoost classifier.
Higher SHAP values indicate greater contribution of temporal-QBER features
to attack classification.}
\label{fig:shap}
\end{figure}

To improve the interpretability of the proposed XGBoost classifier, SHAP\cite{lundberg2018consistent} (SHapley Additive exPlanations) analysis was performed. Fig.~\ref{fig:shap} presents the global mean absolute SHAP importance of the extracted temporal-QBER features. Key generation rate, QBER stability, diagonal-basis QBER, error-gap coefficient of variation, and basis QBER difference are identified as the most influential features. These results indicate that physics-informed temporal and basis-dependent characteristics provide greater discriminative capability than conventional aggregate-QBER features for detecting stealthy eavesdropping attacks.

\subsection{Robustness and Stability Analysis}

\begin{figure}
\centering
\includegraphics[width=0.95\columnwidth]{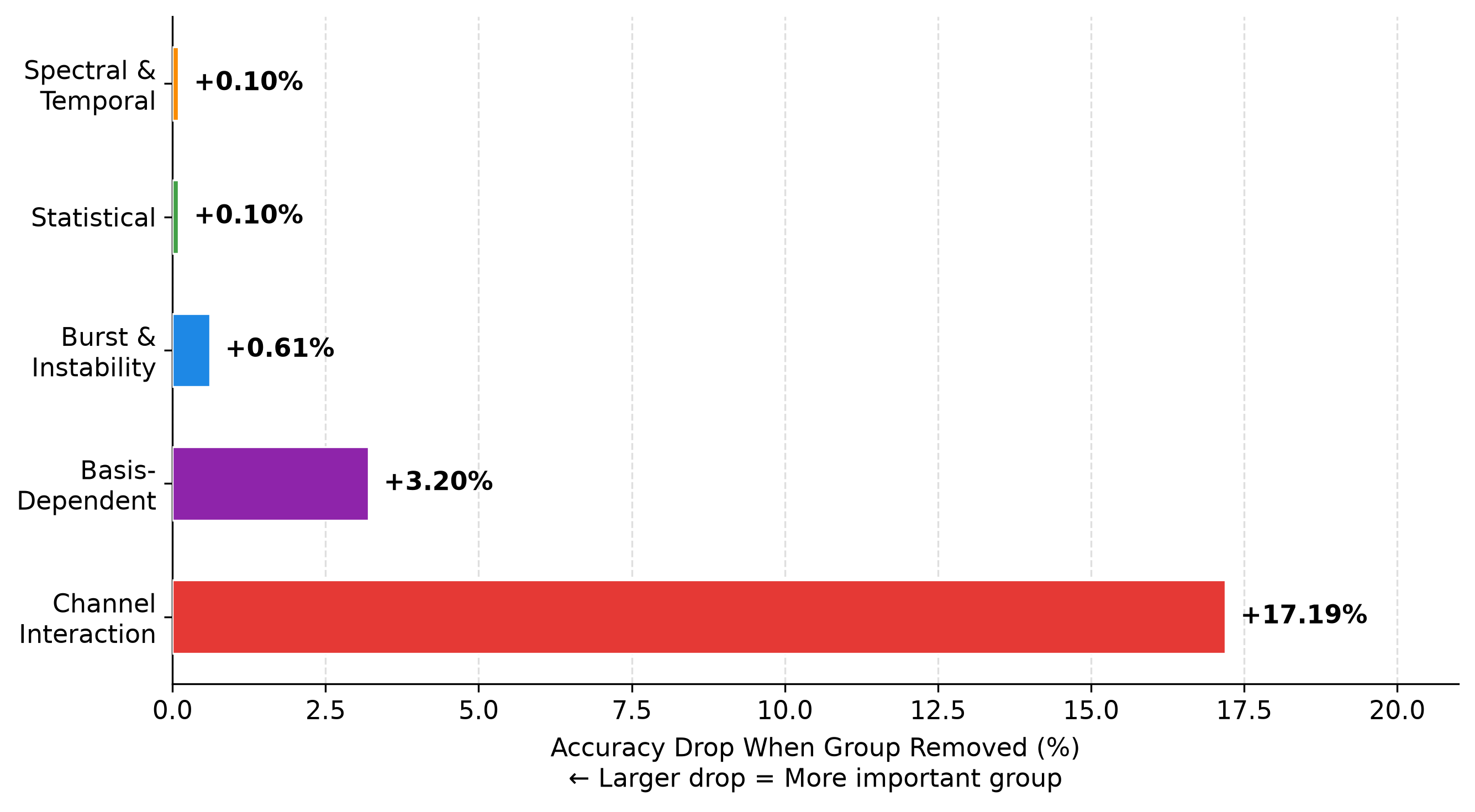}
\caption{Leave-one-group-out feature ablation showing the accuracy reduction after removing each feature group.}
\label{fig:ablation}
\end{figure}

XGBoost demonstrates consistent performance across ten independent runs, achieving an average accuracy of 88.01\% ($\pm$0.47\%) and a macro-F1 score of 0.8803 ($\pm$0.0047). Furthermore, 5-fold stratified cross-validation yields a macro-F1 score of 0.8778 ($\pm$0.0037), indicating minimal overfitting. A leave-one-group-out ablation study Fig.~\ref{fig:ablation} shows that removing the Channel Interaction features causes the largest accuracy reduction (17.19\%), followed by the Basis-Dependent features (3.20\%). This confirms that channel-aware and basis-dependent features contribute most significantly to the proposed detection framework.

\section{DISCUSSION AND FUTURE WORK}

The proposed temporal QBER based framework substantially improves the detection of stealthy eavesdropping attacks compared with conventional threshold-based monitoring while providing accurate and interpretable multi-class attack classification using only standard BB84 post-processing information. The SHAP analysis and feature-group ablation study demonstrate that the proposed physics-informed temporal, basis-dependent, and channel-aware features capture discriminative characteristics that are not available from conventional scalar-QBER monitoring.
Despite these promising results, the present study is limited to simulated BB84 environments with balanced class distributions and simplified attack models. Future work will focus on experimental validation using practical QKD systems, evaluation under class-imbalanced and domain-shift scenarios, investigation of advanced temporal learning approaches, sensitivity analysis of temporal window parameters, and the development of online adaptive attack detection for real-time QKD security monitoring.

\section{Conclusion}

This work presented a temporal-QBER-based machine learning framework for detecting and classifying stealthy eavesdropping attacks in BB84 QKD systems. Using 63 physics-informed features, XGBoost achieved an average accuracy of 88.01\% ($\pm$0.47\%) and a macro-F1 score of 0.8803 across ten independent runs, while reducing the FNR from 0.8477 to 0.0198 compared with the conventional 11\% QBER threshold detector. SHAP-based explainability showed that temporal and channel-aware features provide the strongest discriminative capability for stealthy attack detection. These results demonstrate that the proposed temporal-QBER framework provides an accurate, interpretable, and practical solution for security monitoring in BB84 QKD systems.

\bibliographystyle{IEEEtran}
\bibliography{ref}

@article{bennett1984quantum,
  title={Quantum cryptography: Public key distribution and coin tossing},
  author={Bennett, Charles H and Brassard, Gilles},
  journal={Theoretical Computer Science},
  volume={560},
  pages={7--11},
  year={2014},
  publisher={Elsevier}
}

@article{scarani2009security,
  title={The security of practical quantum key distribution},
  author={Scarani, Valerio and Bechmann-Pasquinucci, Helle and Cerf, Nicolas J 
          and Du{\v{s}}ek, Miloslav and L{\"u}tkenhaus, Norbert and Peev, Momtchil},
  journal={Reviews of Modern Physics},
  volume={81},
  number={3},
  pages={1301--1350},
  year={2009},
  publisher={APS}
}

@article{shor2000simple,
  title={Simple proof of security of the {BB84} quantum key distribution protocol},
  author={Shor, Peter W and Preskill, John},
  journal={Physical Review Letters},
  volume={85},
  number={2},
  pages={441},
  year={2000},
  publisher={APS}
}

@article{lutkenhaus2000security,
  title={Security against individual attacks for realistic quantum key distribution},
  author={L{\"u}tkenhaus, Norbert},
  journal={Physical Review A},
  volume={61},
  number={5},
  pages={052304},
  year={2000},
  publisher={APS}
}

@article{diamanti2016practical,
  title={Practical challenges in quantum key distribution},
  author={Diamanti, Eleni and Lo, Hoi-Kwong and Qi, Bing and Yuan, Zhiliang},
  journal={npj Quantum Information},
  volume={2},
  number={1},
  pages={16025},
  year={2016},
  publisher={Nature Publishing Group}
}

@article{xu2020secure,
  title={Secure quantum key distribution with realistic devices},
  author={Xu, Feihu and Ma, Xiongfeng and Zhang, Qiang and Lo, Hoi-Kwong and Pan, Jian-Wei},
  journal={Reviews of modern physics},
  volume={92},
  number={2},
  pages={025002},
  year={2020},
  publisher={APS}
}

@article{pirandola2020advances,
  title={Advances in quantum cryptography},
  author={Pirandola, Stefano and Andersen, Ulrik L and Banchi, Leonardo and Berta, Mario and Bunandar, Darius and Colbeck, Roger and Englund, Dirk and Gehring, Tobias and Lupo, Cosmo and Ottaviani, Carlo and others},
  journal={Advances in optics and photonics},
  volume={12},
  number={4},
  pages={1012--1236},
  year={2020},
  publisher={Optical Society of America}
}

@book{nielsen2010quantum,
  title={Quantum computation and quantum information},
  author={Nielsen, Michael A and Chuang, Isaac L},
  year={2010},
  publisher={Cambridge university press}
}

@article{brassard2000limitations,
  title={Limitations on practical quantum cryptography},
  author={Brassard, Gilles and L{\"u}tkenhaus, Norbert and Mor, Tal and Sanders, Barry C},
  journal={Physical review letters},
  volume={85},
  number={6},
  pages={1330},
  year={2000},
  publisher={APS}
}

@article{cao2022evolution,
  title={The evolution of quantum key distribution networks: On the road to the qinternet},
  author={Cao, Yuan and Zhao, Yongli and Wang, Qin and Zhang, Jie and Ng, Soon Xin and Hanzo, Lajos},
  journal={IEEE Communications Surveys \& Tutorials},
  volume={24},
  number={2},
  pages={839--894},
  year={2022},
  publisher={IEEE}
}

@article{mehic2020quantum,
  title={Quantum key distribution: a networking perspective},
  author={Mehic, Miralem and Niemiec, Marcin and Rass, Stefan and Ma, Jiajun and Peev, Momtchil and Aguado, Alejandro and Martin, Vicente and Schauer, Stefan and Poppe, Andreas and Pacher, Christoph and others},
  journal={ACM Computing Surveys (CSUR)},
  volume={53},
  number={5},
  pages={1--41},
  year={2020},
  publisher={ACM New York, NY, USA}
}

@article{page1954continuous,
  title={Continuous inspection schemes},
  author={Page, Ewan S},
  journal={Biometrika},
  volume={41},
  number={1/2},
  pages={100--115},
  year={1954},
  publisher={JSTOR}
}

@article{adams2007bayesian,
  title={Bayesian online changepoint detection},
  author={Adams, Ryan Prescott and MacKay, David JC},
  journal={arXiv preprint arXiv:0710.3742},
  year={2007}
}

@article{fanizza2023ultimate,
  title={Ultimate limits for quickest quantum change-point detection},
  author={Fanizza, Marco and Hirche, Christoph and Calsamiglia, John},
  journal={Physical review letters},
  volume={131},
  number={2},
  pages={020602},
  year={2023},
  publisher={APS}
}

@article{ding2023machine,
  title={Machine-learning-based detection for quantum hacking attacks on continuous-variable quantum-key-distribution systems},
  author={Ding, Chao and Wang, Shi and Wang, Yaonan and Wu, Zijie and Sun, Jingtao and Mao, Yiyu},
  journal={Physical Review A},
  volume={107},
  number={6},
  pages={062422},
  year={2023},
  publisher={APS}
}

@inproceedings{tunc2023machine,
  title={Machine learning based attack detection for quantum key distribution},
  author={Tunc, Hilal Sultan Duranoglu and Wang, Yingjian and Bassoli, Riccardo and Fitzek, Frank HP},
  booktitle={2023 IEEE 9th World Forum on Internet of Things (WF-IoT)},
  pages={1--6},
  year={2023},
  organization={IEEE}
}

@article{banerjee2025machine,
  title={Machine Learning assisted noise classification with Quantum Key Distribution protocols},
  author={Banerjee, Shreya and Panigrahi, Prasanta K and others},
  journal={arXiv preprint arXiv:2504.00718},
  year={2025}
}

@article{lizama2016quantum,
  title={Quantum key distribution in the presence of the intercept-resend with faked states attack},
  author={Lizama-P{\'e}rez, Luis Adrian and L{\'o}pez, Jos{\'e} Mauricio and De Carlos L{\'o}pez, Eduardo},
  journal={Entropy},
  volume={19},
  number={1},
  pages={4},
  year={2016},
  publisher={MDPI}
}

@article{al2026resisting,
  title={Resisting quantum key distribution attacks using quantum machine learning},
  author={Al-Kuwari, Ali and Mohamed, Noureldin and Al-Kuwari, Saif and Farouk, Ahmed and Behera, Bikash K},
  journal={IET Quantum Communication},
  volume={7},
  number={1},
  pages={e70028},
  year={2026},
  publisher={Wiley Online Library}
}

@article{al2026machine,
  title={Machine learning techniques for enhancing quantum key distribution},
  author={Al-Kuwari, Ali and Alqrinawi, Safaa and Al-Amir, Lujayn and Mollazehi, Amina and Al-Kuwari, Saif},
  journal={arXiv preprint arXiv:2603.07384},
  year={2026}
}

@inproceedings{saxena2025detection,
  title={Detection Challenges in BB84 Quantum Key Distribution under Random Number Generator Compromise and QBER Analysis},
  author={Saxena, Shashwat and Srivastava, Anand and Bhatia, Vimal and Kumar, Pravindra and Singh, Ritu Raj},
  booktitle={2025 IEEE International Conference on Advanced Networks and Telecommunications Systems (ANTS)},
  pages={1--5},
  year={2025},
  organization={IEEE}
}

@inproceedings{chen2016xgboost,
  title={Xgboost: A scalable tree boosting system},
  author={Chen, Tianqi and Guestrin, Carlos},
  booktitle={Proceedings of the 22nd acm sigkdd international conference on knowledge discovery and data mining},
  pages={785--794},
  year={2016}
}

@article{breiman2001random,
  title={Random forests},
  author={Breiman, Leo},
  journal={Machine learning},
  volume={45},
  number={1},
  pages={5--32},
  year={2001},
  publisher={Springer}
}

@article{lundberg2018consistent,
  title={Consistent individualized feature attribution for tree ensembles},
  author={Lundberg, Scott M and Erion, Gabriel G and Lee, Su-In},
  journal={arXiv preprint arXiv:1802.03888},
  year={2018}
}

@article{cortes1995support,
  title={Support-vector networks},
  author={Cortes, Corinna and Vapnik, Vladimir},
  journal={Machine learning},
  volume={20},
  number={3},
  pages={273--297},
  year={1995},
  publisher={Springer}
}

\end{document}